\documentclass[showpacs,aps,twocolumn]{revtex4}
\usepackage{epsfig}
\usepackage{graphicx}
\usepackage{amsmath,amssymb,amsfonts}
\usepackage{array}
\usepackage{url}
\usepackage{hyperref}
\usepackage{multirow}
\usepackage{float}
\usepackage{lineno}
\usepackage{xspace}
\usepackage[usenames,dvipsnames]{color}

\newcommand{\bef}{\begin{figure}}
\newcommand{\eef}{\end{figure}}
\newcommand{\bc}{\begin{center}}
\newcommand{\ec}{\end{center}}

\newcommand{\be}{\begin{equation}}
\newcommand{\ee}{\end{equation}}
\newcommand{\bea}{\begin{eqnarray}}
\newcommand{\eea}{\end{eqnarray}}

\def\ba{\begin{eqnarray}}
\def\ea{\end{eqnarray}}
\begin{document}
\title{Multiplicity Dependence of Shear Viscosity, Isothermal Compressibility and Speed of Sound in $pp$ collisions at $\sqrt{s}$ = 7 TeV}
\author{Dushmanta Sahu}
\author{Sushanta Tripathy\footnote{Presently at: Instituto de Ciencias Nucleares, UNAM, Deleg. Coyoac\'{a}n, Ciudad de M\'{e}xico 04510}}
\author{Raghunath Sahoo\footnote{Corresponding Author Email: Raghunath.Sahoo@cern.ch}}
\affiliation{Discipline of Physics, School of Basic Sciences, Indian Institute of Technology Indore, Simrol, Indore 453552, India}
\author{Archita Rani Dash}
\affiliation{Department of Physics, School of Advanced Sciences, Vellore Institute of Technology, Vellore 632014 , India}

\begin{abstract}
In order to understand the detailed dynamics of systems produced in $pp$ collisions, it is essential to know about the Equation of State (EoS) and various thermodynamic properties. In this work, we study the shear viscosity to entropy density ratio, isothermal compressibility and speed of sound of the system by considering a differential freeze-out scenario. We have used a thermodynamically consistent Tsallis non-extensive statistics to have a better explanation for the dynamics of $pp$ collision systems. While the shear viscosity to entropy density ratio provides information about the measure of fluidity of a system formed in high energy collisions, the isothermal compressibility gives a clear idea about the deviation of the system from a perfect fluid. The speed of sound in the system as a function of $\langle dN_{\rm ch}/d\eta\rangle$ gives us a vivid picture of the dynamics of the system. The results show quite an intuitive perspective on high multiplicity $pp$ collisions and give us a limit of $\langle dN_{\rm ch}/d\eta\rangle$ $\gtrsim$ (10 - 20), after which a change in the dynamics of the system may be observed.


 \pacs{}
\end{abstract}
\date{\today}
\maketitle{}

\section{Introduction}
\label{intro}
Ultra-relativistic collisions of protons and heavy-ions at the Large Hadron Collider (LHC) at CERN, Switzerland, have been instrumental in  
understanding the sublime nature of the microscopic world at very high energies. One of the most astounding facets of the microscopic realm is a relatively new state of matter called Quark Gluon Plasma (QGP), which is expected to be formed in such collisions. QGP has partons (quarks and gluons) as the degrees of freedom, and exists at very high temperature and/or baryon density. Earlier, it was believed that there wouldn't be any QGP formation in $pp$ collisions. However, recent studies indicate possible formation of QGP droplets in high multiplicity $pp$ collisions \cite{Sahoo:2019ifs,nature,Velicanu:2011zz}. Such reasons compel us to study high energy $pp$ collisions with ever-increasing interest. Understanding the behavior of matter in the hadronic phase of these type of collisions, and having the knowledge about various thermodynamical quantities involved is very useful. The coefficient of shear viscosity to entropy density ratio ($\eta/s$), isothermal compressibility ($\kappa_{\rm T}$) and speed of sound ($c_{\rm s}$) are such thermodynamic quantities that tell us about the fascinating behaviours of the system. 

The space-time evolution of a system formed in ultra-relativistic high energy collisions is governed by its dissipative properties such as the shear viscosity ($\eta$). To understand the behavior of a system formed in such collisions, $\eta/s$ is one of the most important properties. The elliptic flow measurements from heavy-ion collisions at RHIC \cite{STAR-NPA} have found that the medium formed in such collisions gives $\eta /s$ value closer to the KSS bound (Kovtun-Son-Starinets) \cite{Kovtun:2004de}, which might suggest that QGP is almost a perfect fluid \cite{Romatschke:2007mq,Hirano:2005wx}. Thus $\eta/s$ can be used to study the measure of fluidity of a system.

$\kappa_{\rm T}$ gives us information about how the volume of a system changes with the change in pressure at constant temperature \cite{Mukherjee:2017elm}. It tells us about the deviation of a real fluid from a perfect fluid. For a perfect fluid, $\kappa_{\rm T}$ = 0; which means the fluid is incompressible. However, no such fluid exists in nature. But, recent findings have shown that QGP behaves as a nearly perfect fluid, having the lowest $\kappa_{\rm T}$ estimated till now \cite{Sahu:2020nbu}. This also complements previous findings of the ratio of shear viscosity and entropy ($\eta /s$) from ADS/CFT calculations which gives a lower bound (KSS bound) to the ratio \cite{Kovtun:2004de}. 


The study of speed of sound will help us to have a proper idea about the equation of state (EOS) of the system. It plays a crucial role in the hydrodynamical evolution of the matter created in the collisions. It also affects the momentum distributions of the particles created in the collision systems. Observations from heavy-ion collisions have proved that the speed of sound is different in three different phases in the evolution of the collision systems, namely the QGP phase, the mixed phase and the hadronic phase. For a massless non-interacting gas, the value of the squared speed of sound, $c_{\rm s}^{2}$ is 1/3, whereas for a hadron gas the value is around 1/5 \cite{Mohanty:2003va}. The expansion time scale of the system is a measure of the speed of sound which is given by $\tau_{\rm exp}^{-1} \sim \frac{1}{\epsilon}\frac{\partial \epsilon}{\partial \tau_p} = \frac{1 + c_{\rm s}^{2}}{\tau_p}$ \cite{Bjorken}. Here, $\epsilon$ is the energy density of the system and $\tau_p$ is the proper time. While the collision time scale is given by $\tau_{\rm coll}^{-1} \sim n\sigma v$, where $n$ is density of particles, $\sigma$ is the collision cross section and $v$ is the particle velocity. For a system to be in thermal equilibrium, the expansion time scale must be greater than the collision time scale. However, we are taking a differential freeze-out scenario, so it will be interesting to see how the speed of sound varies for different hadronic species as a function of charged particle multiplicity.

A large number of particles are produced in high energy collisions, which demands us to take a statistical approach to study the QCD matter, the particle production and the thermodynamics of the systems. The transverse momentum ($p_{\rm T}$) of the final state particles produced in high energy collisions are expected to follow a thermalized Boltzmann-Gibbs (BG) distribution. However, it is experimentally observed that the $p_{\rm T}$-spectra in $pp$ collisions at the RHIC \cite{Abelev:2006cs,Adare:2011vy} and LHC \cite{Aamodt:2011zj,Abelev:2012cn,Abelev:2012jp,Chatrchyan:2012qb} energies show a
deviation from thermalized Boltzmann distribution. Higher contribution of pQCD effects are responsible for this
deviation and the spectra are better described by a combination of Boltzmann-type exponential and pQCD inspired 
power-law distribution. Although a first principle derivation of Tsallis non-extensive distribution \cite{Tsallis:52} is still a question,
empirically it has been very successful in describing the $p_{\rm T}$-spectra in hadronic collisions. This is used for
getting the multiplicity and thus particle ratios in experimental papers. There are vast theoretical developments 
in this front to bring up the physics messages, which include considering systems formed in hadronic collisions as away from equilibrium, system thermodynamics etc. Although there are various forms of the Tsallis distribution
function used in the literature, a thermodynamically consistent distribution function is used in this paper to describe
the $p_{\rm T}$-spectra in LHC $pp$ collisions \cite{Cleymans:2011in}. The deviation from equilibrium is denoted by a parameter $q$, with $q$ = 1 denoting the equilibrium condition (BG scenario).  At high charged particle multiplicity in high energy collisions, $q$ tends to 1, which is an indication that the system has attained global equilibrium. The extracted thermodynamic parameters \cite{Tsallis:2003vv} such as temperature, $T$ and non-extensivity parameter, $q$ are obtained for different multiplicity classes, which are then used to have the estimation of 
shear viscosity to entropy density ratio, isothermal compressibility and the
speed of sound in the medium (related to the equation of state).

In this paper, we study the shear viscosity to entropy density ratio, the isothermal compressibility and speed of sound in high energy $pp$ collisions. In section \ref{formulation}, we give a brief formulation for shear viscosity, isothermal compressibility and $c_{\rm s}^{2}$ using non-extensive statistics. In section \ref{res}, the results and discussions are given. Finally we summarize our findings in section \ref{sum}.

\section{Formulation}
\label{formulation}
As discussed in the previous section, a finite degree of deviation from the equilibrium statistical description of transverse momentum spectra has been observed by experiments at RHIC and LHC~\cite{Abelev:2006cs,Adare:2011vy,Aamodt:2011zj,Abelev:2012cn,Abelev:2012jp,Chatrchyan:2012qb,Bhattacharyya:2015hya}. In addition, the matter produced in these experiments evolves rapidly in a non-homogeneous way.  Hence, the spatial configuration becomes non-uniform and the global equilibrium in the system is not established~\cite{Randrup:2009gp,Palhares:2009tf,Skokov:2008zp,Skokov:2009yu}. Due to this, some thermodynamic observables become non-extensive. This may also happen if there are local temperature fluctuation and long-range correlations in the produced system~\cite{Wilk:1999dr}. Thus, for the calculation of shear viscosity, we use the relativistic non-extensive Boltzmann transport equation where we assume that a non-equilibrium system, which dissipates energy and  produces entropy relaxes to a local $q$-equilibrium after a certain relaxation time. For a detailed description and calculation, one can follow the Refs.~\cite{Biro:2011bq,Kakati:2017xvr,Kadam:2015xsa}. 
Here, we briefly describe the formulation.

The Boltzmann transport equation (BTE) is given by,
\begin{equation}
\label{eq1}
\frac{\partial f_{\rm p}}{\partial t}+v_{\rm p}^{i}\frac{\partial f_{\rm p}}{\partial x^{i}}+F_{\rm p}^{i}\frac{\partial f_{\rm p}}{\partial p^{i}}=I(f_{\rm p}),
\end{equation}
where $v_{\rm p}^{i}$ is the velocity of the $i$th particle and $F_{\rm p}^{i}$ is the external force acting on that particle, whereas $I(f_{\rm p})$ is the collision integral which gives the rate of change of the non-equilibrium distribution function $f_{\rm p}$ when the system approaches a $q$-equilibrium.

Under the approximation of no external force and proceeding with the relaxation time approximation (RTA), the collision integral can be approximated as,
\begin{equation}
\label{eq2}
I(f_{\rm p})\simeq -\frac{(f_{\rm p}-f_{\rm p}^{0})}{\tau(E_{\rm p})},
\end{equation}
where $\tau(E_{\rm p})$ is called the relaxation time or collision time, which can be interpreted as the mean time between collisions. Also, we consider the thermodynamically consistent Tsallis distribution function, having $q$ value close to 1, as an asymptotic equilibrium function, which is called as a $q$-equilibrium. Thus, Tsallis distribution function is taken as $f_{\rm p}^{0}$ near the local rest frame of the fluid.  One should note here that the above RTA is made in order to simplify the collision term. However, the exact collision term calculation in the non-extensive BTE is more complicated than in the classical BTE and it is beyond the scope of this manuscript. Nevertheless, these calculations can be used to make reasonable first estimations for different transport coefficients~\cite{Biro:2011bq}.

The thermodynamically consistent Tsallis distribution function is given as,
\begin{equation}
\label{eq3}
f_{\rm p}^{0} = \frac{1}{\bigg [1 + (q-1)\bigg(\frac{E_{\rm p} - \bf{p.u} - \mu}{T}\bigg) \bigg]^\frac{q}{q-1}},
\end{equation}
where $\bf{u}$ is the fluid velocity, T and $\mu$ are the temperature and the chemical potential respectively.

The stress-energy tensor is  given by,
\begin{equation}
\label{eq4}
T^{\mu \nu}=T^{\mu \nu}_{\rm 0}+T^{\mu \nu}_{dissipative},
\end{equation}
where $T^{\mu \nu}_{\rm 0}$ is the ideal part and $T^{\mu \nu}_{dissipative}$ is the dissipative part of the stress-energy tensor. When we provide hydrodynamical description of QCD, the shear viscosity ($\eta$) and bulk viscosity ($\zeta$) are included in the dissipative part of stress-energy tensor, which can be written in the local Lorentz frame as,
\begin{equation}
\label{eq5}
T^{ij}= -\eta \bigg(\frac{\partial u^{i}}{\partial x^{j}}+\frac{\partial u^{j}}{\partial x^{i}} \bigg)-\bigg(\zeta -\frac{2}{3}\eta \bigg)\frac{\partial u^{i}}{\partial x^{j}}\delta^{ij}.
\end{equation}

In terms of distribution function, the above expression becomes,
\begin{equation}
\label{eq6}
T^{ij}=\int_{}^{}\frac{d^{3}p}{(2\pi)^{3}}\frac{p^{i}p^{j}}{E_{\rm p}}\delta f_{\rm p}.
\end{equation}
Here $\delta f_{\rm p}$ is the deviation of the distribution function from the $q$-equilibrium and is given by (from Eq. \ref{eq1} and \ref{eq2}),
\begin{equation}
\label{eq7}
\delta f_{\rm p}= -\tau(E_{\rm p})\bigg( \frac{\partial f_{\rm p}^{0}}{\partial t}+v_{\rm p}^{i}\frac{\partial f_{\rm p}^{0}}{\partial x^{i}} \bigg).
\end{equation}

Under the assumption of a steady flow of the form $u^{i} = (u_{\rm x}(y),0,0)$ and space-time independent temperature, Eq. \ref{eq5} becomes $T^{xy} = -\eta \partial u_{\rm x}/\partial y$.
From Eq. \ref{eq6} and \ref{eq7}, by using $\mu$ = 0 (for LHC energies), we get,
\begin{equation}
\label{eq8}
T^{xy}=\bigg\{ -\frac{1}{T} \int_{}^{} \frac{d^{3}p}{(2\pi)^{3}}\tau(E_{\rm p})\bigg(\frac{p_{\rm x}p_{\rm y}}{E_{\rm p}} \bigg)^{2}q~(f_{\rm p}^{0})^{\frac{2q-1}{q}}\bigg\}\frac{\partial u_{\rm x}}{\partial y}.
\end{equation}

For a single component of hadronic matter, the coefficient of shear viscosity $\eta$ can be expressed with the non-extensive parameters as,
\begin{equation}
\label{eq9}
\eta = \frac{1}{15T}\int_{ }^{ }\frac{d^{3}p}{(2\pi)^{3}}\tau(E_{\rm p})\frac{p^{4}q}{E_{p}^{2}}(f_{\rm p}^{0})^{\frac{(2q-1)}{q}}
\end{equation}

The energy dependent relaxation time is given by,
\begin{equation}
\label{eq10}
\tau^{-1}(E_{a})=\sum_{bcd}\int_{}^{}\frac{d^{3}p_{b}d^{3}p_{c}d^{3}p_{d}}{(2\pi)^{3}(2\pi)^{3}(2\pi)^{3}}W(a,b\to c,d) f_{b}^{0},
\end{equation}
where $E_{a}$ is the energy of the $a$th particle and $f_{b}^{0}$ is the distribution function for $b$th particle. $W(a,b\to~c,d)$ is the transition rate defined as,
\begin{equation}
\label{eq11}
W(a,b\to c,d)=\frac{2\pi^{4}\delta(p_{a}+p_{b}-p_{c}-p_{d})}{2E_{a}2E_{b}2E_{c}2E_{d}}|\mathcal{M}|^{2},
\end{equation}
where $|\mathcal{M}|$ is the transition amplitude. In the centre-of-mass frame Eq. \ref{eq10} can be written as,
\begin{eqnarray}
\label{eq12}
\tau^{-1}(E_{a})=\sum_{b}\int_{}^{}\frac{d^{3}p_{b}}{(2\pi)^{3}}\sigma_{ab}\frac{\sqrt{s-4m^{2}}}{2E_{a}2E_{b}}f_{b}^{0}\nonumber\\
\equiv \sum_{b}\int_{}^{}\frac{d^{3}p_{b}}{(2\pi)^{3}}\sigma_{ab}v_{ab}f_{b}^{0}
\end{eqnarray}
where $v_{ab}$ is the relative velocity and $\sqrt{s}$ is the centre-of-mass energy and $\sigma_{ab}$ is the total scattering cross-section in the process. $\tau(E_{a})$ can be further written approximately as the averaged relaxation time ($\tilde{\tau}$), which can be obtained from Eq. \ref{eq12} by averaging over $f_{a}^{0}$ as,
\begin{equation}
\label{eqy}
\tilde{\tau_{a}}^{-1}=\frac{\int_{}^{}\frac{d^{3}p_{a}}{(2\pi)^{3}}\tau^{-1}(E_{a})f_{a}^{0}}{\int_{}^{}\frac{d^{3}p_{a}}{(2\pi)^{3}}f_{a}^{0}}\nonumber\\
= \sum_{b}\frac{\int_{}^{}\frac{d^{3}p_{a}}{(2\pi)^{3}}\frac{d^{3}p_{b}}{(2\pi)^{3}}\sigma_{ab}v_{ab}f_{a}^{0}f_{b}^{0}}{\int_{}^{}\frac{d^{3}p_{a}}{(2\pi)^{3}}f_{a}^{0}}\nonumber
\end{equation}
\begin{equation}
\label{eq13}
=\sum_{b}n_{b}\langle\sigma_{ab}v_{ab} \rangle.
\end{equation}
$n_{b}$ is the number  density of $b$th particle. Here, all the calculations are done for one single hadronic species at a time. The thermal average of the scattering of the same species of particles with constant cross-section and zero baryon density can be estimated as,
\begin{equation}
\label{eq14}
\langle\sigma_{ab}v_{ab} \rangle=\frac{\sigma \int_{}^{}d^{3}p_{a}d^{3}p_{b}v_{ab}e_{q}^{-\frac{E_{a}}{T}}e_{q}^{-\frac{E_{b}}{T}}}{\int_{}^{}d^{3}p_{a}d^{3}p_{b}e_{q}^{-\frac{E_{a}}{T}}e_{q}^{-\frac{E_{b}}{T}}},
\end{equation}
where $e_{q}^{(x)}$ is the $q$-exponential which is defined as,
\begin{equation}
\label{eq15}
e_{q}^{(x)}=[1+(q-1)x]^{q/(q-1)}.
\end{equation}
The momentum space volume elements are be written as,
\begin{equation}
\label{eq16}
d^{3}p_{a}d^{3}p_{b}=8\pi^{2}p_{a}p_{b}dE_{a}dE_{b}~dcos\theta.
\end{equation}

On solving, Eq. \ref{eq14} can be further written as,
\begin{widetext}
\begin{equation}
\label{eq17}
\langle\sigma_{ab}v_{ab} \rangle=\frac{\sigma\int_{}^{}8\pi^{2}p_{a}p_{b}dE_{a}dE_{b}~dcos\theta e_{q}^{-\frac{E_{a}}{T}}e_{q}^{-\frac{E_{b}}{T}}\frac{\sqrt{(E_aE_b-p_{a}p_{b}cos\theta)^{2}-(m_{a}m_{b})^{2}}}{{E_{a}E{_b}-p_{a}p_{b}cos\theta}}     }{\int_{}^{}8\pi^{2}p_{a}p_{b}dE_{a}dE_{b}~dcos\theta e_{q}^{-\frac{E_{a}}{T}}e_{q}^{-\frac{E_b}{T}}}.
\end{equation}
\end{widetext}
$\sigma$ is the hadronic collision cross-section and a constant value of 11.3 mb~\cite{Kakati:2017xvr,Kadam:2015xsa} is used in our calculations.
The thermodynamical quantities in non-extensive statistics are calculated as \cite{Cleymans:2012ya},
\begin{equation}
\label{eq18}
n = g \int \frac{d^{3}p}{(2\pi)^3}\bigg[1 + (q-1)\frac{E- \mu}{T} \bigg]^\frac{-q}{q-1}
\end{equation}
\begin{equation}
\label{eq19}
\epsilon = g \int \frac{d^{3}p}{(2\pi)^3} E \bigg[1 + (q-1)\frac{E- \mu}{T} \bigg]^\frac{-q}{q-1}
\end{equation}
\begin{equation}
\label{eq20}
P = g \int \frac{d^{3}p}{(2\pi)^3} \frac{p^{2}}{3E} \bigg[1 + (q-1)\frac{E- \mu}{T} \bigg]^\frac{-q}{q-1}.
\end{equation}
$n$, $\epsilon$ and $P$ are the number density, energy density and pressure of hadrons, respectively. $g$ is the particle degeneracy.
Also, the non-extensive entropy density is given by,
\begin{equation}
\label{eq21}
s = \frac {\epsilon + P - \mu n}{T}
\end{equation}
From thermodynamics, the isothermal compressibility ($\kappa_{\rm T}$) is defined as~\cite{KHuang, Landau, Kakati:2017xvr},
\begin{equation}
\label{eq22}
\kappa_{\rm T} = -\frac{1}{V}\frac{\partial V}{\partial P}\bigg\vert_T,
\end{equation}
where V, P and T are the volume, pressure and temperature of the system. In terms of multiplicity fluctuations and average number, isothermal compressibility can be defined as,
\begin{equation}
\label{eq23}
\bigg\langle (N - \langle N \rangle)^{2} \bigg\rangle = var(N) = \frac{T \langle N \rangle^{2}}{V}\kappa_T,
\end{equation}
where $N$ is the particle multiplicity.
From basic thermodynamic relation, we have
\begin{equation}
\label{eq24}
\bigg\langle (N - \langle N \rangle)^{2} \bigg\rangle = VT\frac {\partial n}{\partial \mu}.
\end{equation}
Thus, from Eq.\ref{eq23} and Eq.\ref{eq24} we derive the expression,
\begin{equation}
\label{eq25}
\kappa_{\rm T} = \frac {\partial n/\partial \mu}{n^{2}}
\end{equation}
where,
\begin{equation}
\label{eq26}
\frac {\partial n}{\partial \mu} = \frac{gq}{T} \int \frac{d^{3}p}{(2\pi)^3}\bigg[1 + (q-1)\frac{E- \mu}{T} \bigg]^\frac{1-2q}{q-1}
\end{equation}
At the LHC energies, the baryochemical potential of the system is almost zero and for our studies, we use $\mu$ = 0 in the
calculations. To have a better understanding about a system, it is important to know the Equation of State (EoS) which is given by the speed of sound in that system. Speed of sound squared is given by \cite{Cleymans:2011fx},
\begin{equation}
\label{eq27}
c_{\rm s}^{2} = \bigg (\frac {\partial P}{\partial \epsilon}\bigg)_{\rm s/n}.
\end{equation}
It can be further written as,
\begin{equation}
\label{eq28}
c_{\rm s}^{2} = \frac {(\frac {\partial P}{\partial T})} {(\frac {\partial \epsilon}{\partial T})}.
\end{equation}

\section{Results and Discussion}
\label{res}

\begin{figure}[ht!]
\begin{center}
\includegraphics[scale = 0.48]{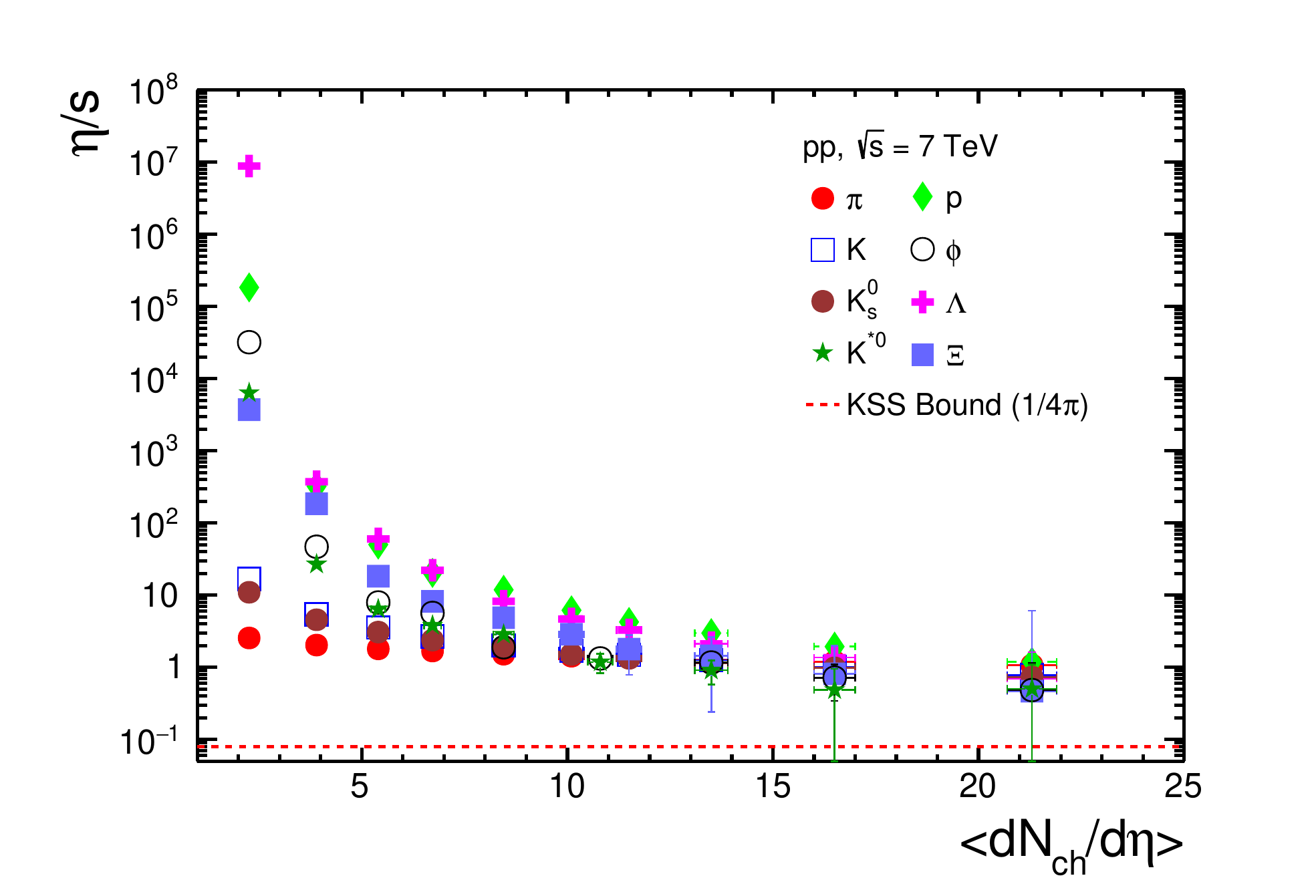}
\caption{(Color online) $\eta/s$ as a function of charged particle multiplicity for $pp$ collisions at $\sqrt{s}$ = 7 TeV for different charged particles.}
\label{fig1}
\end{center}
\end{figure}

The experimental data from ALICE for pp collisions at $\sqrt{s}$ = 7 TeV \cite{Acharya:2018orn} are used for this analysis. We have taken T and $q$ values from Tsallis distribution function by fitting the $p_{\rm T}$-spectra of produced identified particles in a differential freeze-out scenario \cite{Khuntia:2018znt}. The Eq. \ref{eq9} and Eq. \ref{eq21} are used to calculate the shear viscosity to entropy density ratio. Fig.~\ref{fig1} shows the coefficient of shear viscosity to entropy density ratio as a function of charged particle multiplicity. 
The final state charged particle multiplicity density in pseudorapidity, $\langle dN_{\rm ch}/d\eta\rangle$ is used as an event classifier
in $pp$ collisions at the LHC energies in order to classify low-multiplicity and high-multiplicity events -- like centrality classification in 
heavy-ion collisions. This is required to study various observables as a function of final state system multiplicity.
We observe that $\eta/s$ decreases with increase in $\langle dN_{\rm ch}/d\eta\rangle$ and at higher multiplicity it becomes the lowest approaching the KSS bound value. As pion being the lightest particle among the studied particles, the density of pions is higher in a system formed in high energy collisions. Thus, pions exhibit lowest $\eta/s$ at low multiplicity and all the particles approach to a minimum $\eta/s$ at high-multiplicity $pp$ collisions. At high multiplicity beyond $\langle dN_{\rm ch}/d\eta\rangle$ $\gtrsim$ (10 - 20),  $\eta/s$ becomes almost constant and remain the same for all the particles. Thus the limit $\langle dN_{\rm ch}/d\eta\rangle$ $\gtrsim$ (10 - 20) can be interpreted as the threshold after which the system seems to be going through a change in its dynamics. This is one of the significant observations in the current work due to the fact that we use an experimentally motivated and thermodynamically consistent Tsallis distribution to explain the experimental data and the obtained results indicate a possible formation of QGP droplets in high multiplicity $pp$ collisions.

\begin{figure}[ht!]
\begin{center}
\includegraphics[scale = 0.45]{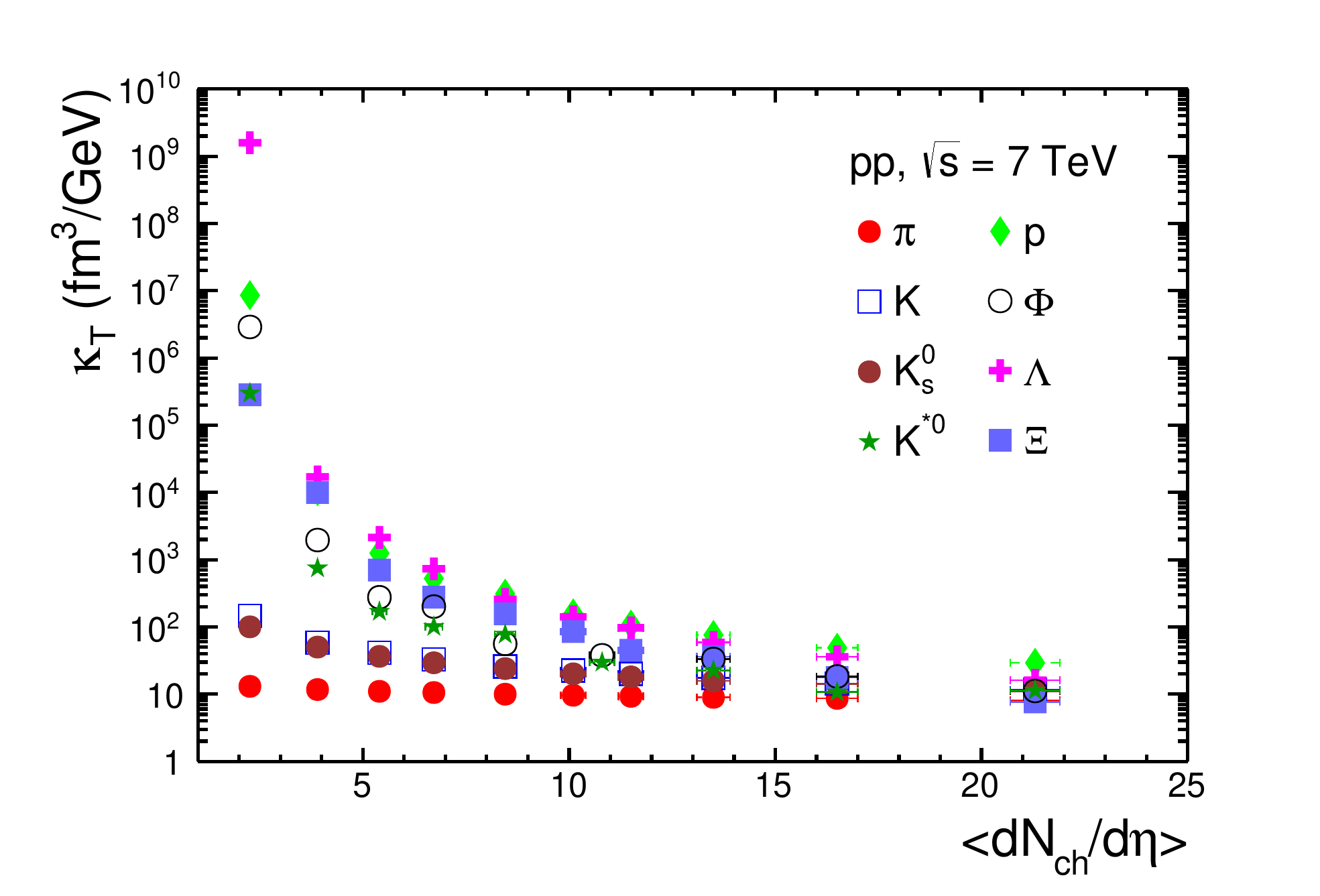}
\caption{(Color online) $\kappa_{\rm T}$ as a function of charged particle multiplicity for $pp$ collisions at $\sqrt{s}$ = 7 TeV for different final state particles.}
\label{fig2}
\end{center}
\end{figure}

In order to look deeper into the possibility of a near perfect fluid behaviour of the produced system, let us now study another important observable in this direction, {\it i.e.} the isothermal compressibility. For this, we take various identified particles produced in $pp$ collisions and use Eq.\ref{eq25} to estimate isothermal compressibility in which T and $q$ values are taken by fitting Tsallis distribution function to the $p_{\rm T}$-spectra \cite{Acharya:2018orn,Khuntia:2018znt}.  Figure \ref{fig2} shows the variation of $\kappa_{\rm T}$ of the identified particles as a function of charged particle multiplicity.
We observe that $\kappa_{\rm T}$ decreases with the increase in $\langle dN_{\rm ch}/d\eta\rangle$. This result is in agreement with our previous findings \cite{Sahu:2020nbu}. For lighter particles,
the values of $\kappa_{\rm T}$ are lower and the values increase as the mass of the particles increase till a certain $\langle dN_{\rm ch}/d\eta\rangle$. Pion being the lightest meson, has the lowest $\kappa_{\rm T}$ as compared to the others. This is because, $\kappa_{\rm T}$ is inversely proportional to the number density of the system. As pion number density is highest in a collision system, its isothermal compressibility is the lowest. However, as it is clearly seen, for higher charged particle multiplicity ($\langle dN_{\rm ch}/d\eta\rangle$ $\gtrsim$ 10 - 20), the $\kappa_{\rm T}$ of all the hadrons converge together and show only a slight variation from each other. Beyond this threshold limit, a QGP-like medium formation is expected regardless of the collision systems \cite{Sahu:2019tch}.  It is worth mentioning here that a differential kinetic freeze-out scenario becomes a single freeze-out as has been discussed in heavy-ion collisions for a multiplicity threshold of $\langle dN_{\rm ch}/d\eta\rangle$ $\gtrsim$ 10 - 20, which is 
one of the important findings of the present study.

The isothermal compressibility of water at room temperature is reported to be 6.62$\times$10$^{42}$ fm$^{3}$/GeV \cite{water}. We have estimated the $\kappa_{\rm T}$ of the hadron gases to be around 10 fm$^{3}$/GeV at high charged particle multiplicity. This value is still larger than the isothermal compressibility found for QGP-like medium in our previous work, which was found to be $\sim$ 0.3 fm$^{3}$/GeV \cite{Sahu:2020nbu}.

\begin{figure}[ht!]
\begin{center}
\includegraphics[scale = 0.45]{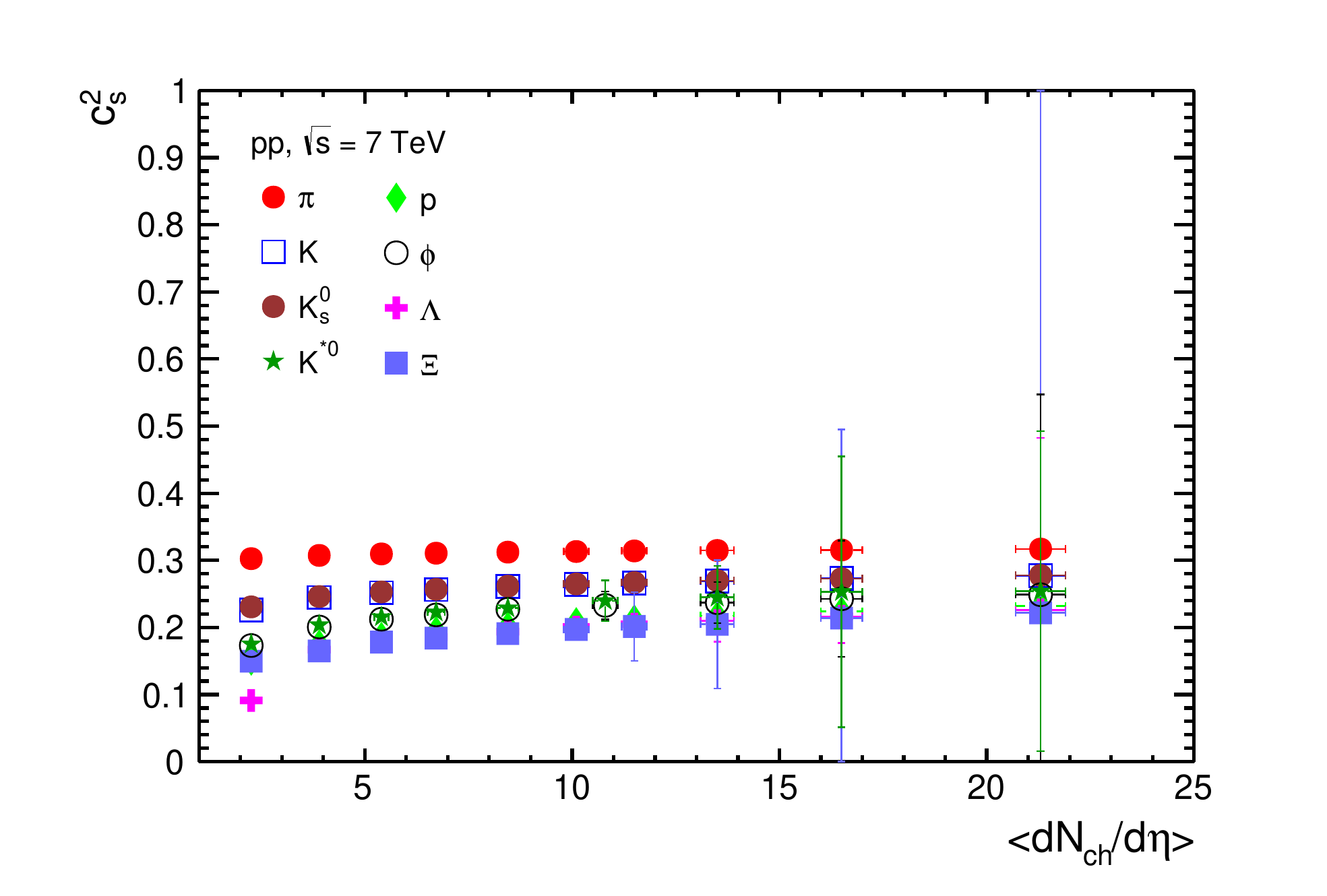}
\caption{(Color online) Speed of sound squared vs charged particle multiplicity for $pp$ collisions at $\sqrt{s}$ = 7 TeV for different 
final state particles.}
\label{fig4}
\end{center}
\end{figure}

The speed of sound squared ($c_{\rm s}^{2}$) being related to the equation of state of the system (EoS), is one of the important thermodynamic observables. In Fig. \ref{fig4}, we have plotted the $c_{\rm s}^{2}$ as a function of charged particle multiplicity, which is calculated using Eq.\ref{eq28}. We observe that $c_{\rm s}^{2}$ increases with the increase in $\langle dN_{\rm ch}/d\eta\rangle$. Here we see a mass ordering between the identified particles. As pion is the most abundant particle in the system, the $c_{\rm s}^{2}$ for pion is the highest. These observations complement the previous estimations \cite{Deb:2019yjo,Khuntia:2016ikm}. The other particles follow the same trend accordingly. At low $\langle dN_{\rm ch}/d\eta\rangle$, the speed of sound is lower. This is because the $c_{\rm s}^{2}$ depends on the density of the system. For low charged particle multiplicity, the density is lower, and it increases slowly with increase in $\langle dN_{\rm ch}/d\eta\rangle$.  However, after a certain charged particle multiplicity, the density of the system doesn't change \cite{Sharma:2018uma}. We observe that there is a transition in the behavior of the $c_{\rm s}^{2}$ plot at around $\langle dN_{\rm ch}/d\eta\rangle$ $\gtrsim$ 10 - 20, and almost becomes constant after this limit. This gives us a hint about a change in the dynamics of the systems after certain charged particle multiplicity. The squared value of the speed of sound in air at room temperature is 1.3$\times$10$^{-12}$ and in distilled water at room temperature it is around 2.5$\times$10$^{-11}$ \cite{uniphy}. With comparison to these, the $c_{\rm s}^{2}$ of the hadron gases we found here is very high. This indicates the fact that, the hadron gas systems formed in high energy collisions are very dense mediums and at higher charged particle multiplicities, the value of $c_{\rm s}^{2}$ tend towards 1/3, which means the medium behaves like almost an ideal gas.

\section{Summary}
\label{sum}
In summary,
\begin{enumerate}
\item We have estimated the shear viscosity to entropy density ratio of the system produced in $pp$ collision at $\sqrt{s}$ = 7 TeV energy and observed that at higher charged particle multiplicity, $\eta/s$ becomes the lowest, approaching the KSS bound. This helps us to conclude that at higher charged particle multiplicities, the system could be non-dissipative in nature.

\item In order to look deeper into the possibility of a near perfect fluid behaviour of the produced system, we have also estimated the isothermal compressibility of the hadron gases in $pp$ collision system by considering differential freeze-out scenario. We observed that $\kappa_{\rm T}$ of the systems decreases with the increase in the charged particle multiplicity. This suggests that at higher charged particle multiplicity, the system is less compressible. 

\item The shear viscosity to entropy density ratio and isothermal compressibility becoming independent of particle species around $\langle dN_{\rm ch}/d\eta \rangle$ $\gtrsim$ 10-20 is an indication of a transition from differential to single kinetic freeze-out, as is observed in heavy-ion collisions. This observed threshold in the final state charged particle density is an important finding in view of the scaling observed in the LHC energies 
across different collision species.


\item The speed of sound squared for different particles are estimated. We see a range of values from 0.15 to 0.32 for different particles, with pion having the highest $c_{\rm s}^{2}$. 

\item In all our findings, we have observed a limit of $\langle dN_{\rm ch}/d\eta\rangle$ $\gtrsim$ 10 - 20 after which the system appears to be going through some change in its dynamics. This limit of charged particle multiplicity may suggest a requirement for the possible formation of QGP droplets in high multiplicity $pp$ collisions \cite{Campanini:2011bj,Sahoo:2018orz}.

\item Although in this paper, a theoretical estimation of various important thermodynamic quantities are made as a function of final state multiplicity in $pp$ collisions in order to carry out a systematic study, the physical interpretation of the low multiplicity system has to be done with caution.
\end{enumerate}
It would be interesting to go to higher multiplicity classes by extending the same analysis to look for possible criticality in the
system, through the inclusion of higher collision energies and collision 
species ($p$-Pb, Pb-Pb). However, it should be noted that higher final state multiplicity will lead the system towards a Boltzmann-Gibbs statistical description. A smooth transition in terms of system dynamics and statistical mechanics 
description from hadronic to nuclear collision has been an issue always.
\section*{Acknowledgement} 
The authors acknowledge the financial supports from ALICE Project No. SR/MF/PS-01/2014-IITI(G) of Department of Science \& Technology, Government of India. R. S. acknowledges the financial supports from DAE-BRNS Project No. 58/14/29/2019-BRNS.

{}

\end{document}